\documentstyle[12pt]{article}  
\begin{document}   
\baselineskip 18pt
\begin{center}  
{\Large Electroweak  Baryogenesis and Constraints
on Left-handed Majorana Neutrino Masses}\\ 
\vskip .5in

{\large    Utpal     Sarkar\footnote{     e-mail     address    :
utpal@prl.ernet.in}\\[.3in]

Theory  Group \\  
Physical  Research  Laboratory  \\  
Ahmedabad - 380009, India.}

\end{center}

\vskip .75in 
\begin{abstract} 
\baselineskip 16pt

The lepton number violating  interactions  generated by the light
Majorana  neutrinos can erase the primordial  baryon asymmetry of
the  universe  during  the  electroweak   phase  transition.  The
Majorana masses of the  left-handed  neutrinos are constrained to
avoid  this  problem.  These  constraints  do not  depend  on the
$(B-L)$  symmetry  breaking  mechanism.

\end{abstract}

\newpage 
\vskip 1.5in
\baselineskip 18pt

The question of  baryogenesis  \cite{olive}  revived  when it was
noticed that the $SU(2)_L$  anomaly  corresponding  to the global
baryon   number   is   non-vanishing   \cite{hooft}   and   these
interactions may be fast enough to wash out the primordial baryon
asymmetry of the universe in the presence of the sphaleron fields
\cite{krs}.  Attempts were then made to generate baryon asymmetry
during       the       electroweak        symmetry       breaking
\cite{ewbar,shapos,mstv}.  Although  this  approach is  extremely
elegant, here one needs to protect the generated baryon asymmetry
after  the  phase   transition.  This  puts   strong   constraint
\cite{higgsbound}  on the mass of the  higgs  doublets.  With the
present  experimental limit on the higgs mass of about 60 GeV, it
is  already  difficult  to save this  scenario.  If the  higgs is
found  to be  heavier  than 80  GeV,  then  one  has to look  for
alternative solution to the baryogenesis problem.

It has  been  shown  that  if  there  is any  primordial  $(B-L)$
asymmetry  that would get converted to a baryon  asymmetry of the
universe  during the  electroweak  phase  transition  whereas the
$(B+L)$  asymmetry  will be washed out.  Lepton number  violating
$CP$  nonconserving  decays of the heavy Majorana  neutrinos were
then found to generate enough $(B-L)$ asymmetry, which could then
get  converted to baryon  asymmetry  of the  universe  during the
electroweak phase transition \cite{fy1}.

This makes the study of lepton number violating interaction quite
important before the electroweak  symmetry breaking.  If there is
fast lepton  number  violation  before the  electroweak  symmetry
breaking is over, then that would wash out any $(B-L)$ asymmetry.
Since  $(B+L)$  asymmetry  is also  washed  out by the  anomalous
processes,  there will not be any baryon asymmetry left after the
electroweak  phase  transition.  It has been pointed out that the
condition that any lepton number violating interactions (such as,
$ l_L + \phi^\dagger  \rightarrow l_L^c + \phi$ through the right
handed  neutrino $N$ exchange or the lepton number  violating $N$
decays)  should  not be  faster  than the  expansion  rate of the
universe  constraints  the mass  ($M_N$)  of the heavy  neutrinos
\cite{nb1,nb2}.  Here  $l_L$  are the  usual  left-handed  lepton
doublets  and $\phi$ is the doublet  higgs  scalar  field,  which
breaks the electroweak  symmetry at the scale $M_W$.  It was then
argued that since, in many  models,  the  Majorana  masses of the
left-handed  leptons  are related to the  Majorana  masses of the
right  handed  neutrinos,  they relate this bound on $M_N$ to the
Majorana masses of the left-handed neutrinos  \cite{nb1,nb2}.  If
the heavy Majorana neutrinos couple to the left-handed  neutrinos
through a Yukawa  coupling which gives a Dirac mass matrix $m_D$,
then the Majorana mass matrix of the left-handed neutrinos can be
given  by, $$ m_\nu = m_D^T {1 \over  M_N} m_D .  $$ This is true
for the  see-saw  mechanism  of  neutrino  masses  \cite{seesaw}.
Although see-saw mechanism of generating light neutrinos is quite
natural in the left-right  symmetric  models \cite{lr} and is the
most popular model, there exists several other models where light
neutrinos   are   not   generated    using   see-saw    mechanism
\cite{e6,zee,tech}.

In this note I shall first argue that  existing the bounds on the
left-handed neutrinos from baryogenesis are model dependent (even
in the  framework  of see-saw  mechanism)  and there are  several
cases,  where  these  bounds are not  relevant.  Then I point out
that  the  Majorana  masses  of  the  left-handed  neutrinos  can
generate lepton number violating processes during the electroweak
phase transition, which can constraint the Majorana masses of the
left-handed neutrinos directly.

It  has  already  been   mentioned  in  the  earlier   references
\cite{nb2}   that  the  bound  on  the  Majorana  masses  of  the
left-handed  neutrinos  obtained from an analysis of the bound on
the the heavy right handed neutrinos are not valid if some global
U(1) symmetry is exactly conserved up to an electroweak  anomaly.
In addition it is clear that if the left-handed  neutrino mass is
not related to any heavy  neutrinos  through  see-saw  mechanism,
then  also  these  bounds on the  left-handed  neutrinos  are not
valid.  A  large   class  of   models   fall  in  this   category
\cite{e6,zee}.  If the  determinant  of the heavy  neutrino  mass
matrix vanishes, in that case also one cannot apply the bounds on
the left-handed neutrinos.

Over and above these  general  cases, there are several  specific
cases even  within the  framework  of see-saw  models,  where the
bound  given from an  analysis  of the bound on the right  handed
neutrino  fails.  Consider,  for  example, a scenario  with three
generations  of right handed  neutrinos  $N_i$  ($i=1,2,3$).  For
simplicity we assume the Majorana mass matrix for these particles
is diagonal  and real $M_N = $ diag $(M_1, M_2,  M_3)$.  The part
of the  lagrangian  which  controls  the neutrino  mass matrix is
given by,

$$  {\cal  L} =  M_{\alpha}  N_\alpha  N_\alpha  +  h_{\alpha  i}
\overline{N_\alpha} l_{L i} \phi + h.c.  $$

\noindent where $l_{L i}$ are the left-handed lepton doublets ($i
=  1,2,3$  is the  generation  index)  and  $\phi$  is the  higgs
doublets  of the  standard  model  which  breaks the  electroweak
symmetry.  We also assume the hierarchy  $M_1 < M_2 < M_3$, which
means  that  after the  decays of the the $N_2$ and  $N_3$ we are
left with only $N_1$ at the mass scale around $M_1$.  Whether the
decays of $N_2$ or $N_3$ has  generated  any lepton  asymmetry or
not, if the $N_1$ decays at  equilibrium  then this will wash out
all lepton  asymmetry  (in which case one can still have  $(B-L)$
asymmetry generated through $\Delta B = 2$ transitions  \cite{mm}
at a later  stage).  On the other  hand if $N_1$  decays  are not
very fast so as not to erase any lepton asymmetry or it generates
any  lepton  asymmetry  the final  amount of  asymmetry  will not
depend on how fast $N_2$ and $N_3$ decayed.  Hence when any bound
on the mass of the heavy  neutrinos are discussed it is the bound
on $M_1$.

Now consider the condition that the decay rate of $N_1$ is slower
than the expansion rate of the universe,

\begin{equation}
\frac{\sum_i |h_{1 i}^\dagger h_{1 i}|}{16 \pi} M_1 < H = 
1.7 \sqrt{g_*} \frac{T^2}{M_p} \hskip 1in {\rm at} \:\:\: T = M_1
\end{equation}

\noindent This gives a bound on the quantity  $\frac{\sum_i |h_{1
i}^\dagger  h_{1  i}|}{M_1}$.  However, the physical  left-handed
neutrino masses will depend on the other components of $h_{\alpha
i}$ and $M_N$ also.

Consider a very simple example where the matrix $h_{\alpha
i}$ is also diagonal, and hence the Dirac mass matrix $
m_{ \alpha i}^D = h_{\alpha i} \langle \phi \rangle $ is 
given by, 
\begin{equation}
m_{ \alpha i}^D = \pmatrix{m_{1} & 0 & 0 \cr
0 & m_{2} & 0 \cr
0 & 0 & m_{3} } .
\end{equation}
The baryogenesis bound reads, $ \frac{  {(m^{D}_{11})}^2}{M_1}  <
10^{-3}$  eV (which is  satisfied  in most  models),  whereas the
neutrino masses for the second and third generation are, $ \frac{
{(m^{D}_{22})}^2}{M_2}$  and  $  \frac{   {(m^{D}_{33})}^2}{M_3}$
respectively, which are not constrained.

With the  above  arguments  it is clear  that the  bounds  on the
Majorana  masses  of  the  left-handed   neutrinos,  which  comes
indirectly  from the bound on the  Majorana  masses  of the right
handed  neutrinos,  can be avoided in many models.  We now try to
discuss the question if one can  constraint  the Majorana mass of
the  left-handed  neutrinos  directly.  This  is  possible  since
during the  electroweak  phase  transition,  as soon as the higgs
doublets  acquires a vacuum  expectation  value  ($vev$)  and the
$SU(2)_L$ group is broken, there is no symmetry which can prevent
the mass of the  left-handed  neutrinos.  So if lepton  number is
broken before the  electroweak  symmetry,  then as soon as $\phi$
acquires  a  $vev$  the  left-handed  neutrinos  will  get a mass
$(m_\nu)$.  This  can,  in  principle,   induce   lepton   number
violating  processes very fast, which can wash out any primordial
$(B-L)$  asymmetry, which in addition to the sphaleron  processes
will wash out all baryon asymmetry of the universe.

Hence any lepton number violating processes due to the Majorana
mass of the light neutrinos ($m_\nu$) should be slower than the expansion 
rate of the universe, otherwise that will wash out any $(B-L)$ 
asymmetry of the universe and hence the baryon asymmetry during
the electroweak phase transition. There will be several lepton
number violating processes, which will be active at that time.
Consider for example, the process,
\begin{equation}
W^+ + W^+ \rightarrow e^+_i + e^+_j \hskip .5in {\rm and}
\hskip .5in W^- + W^- \rightarrow e^-_i + e^-_j
\end{equation}
mediated by a virtual left-handed neutrino exchange as shown
in figure 1. Here $i$ and $j$ are the generation indices. Depending
on the physical mass (and also on the elements of the mass matrix)
of the left-handed Majorana neutrinos these processes can wash
out any baryon asymmetry between the time when the higgs
acquires a $vev$ and the $W^\pm$ decays out, {\it i.e.}, between
the energy scales 250 GeV and 80 GeV. The condition that these 
processes will be slower than the expansion rate of the universe,
\begin{equation}
\Gamma (WW \to e_i e_j) = \frac{\alpha_W^2 
{(m_\nu)}^2_{ij} T^3}{ m_W^4} < 1.7 \sqrt{g_*}
\frac{T^2}{M_p} \hskip 1in {\rm at} \:\:\: T = M_W
\end{equation}
gives a bound on the Majorana mass of the left-handed neutrinos,
\begin{equation}
{(m_\nu)}^2_{ij} < 20 keV .
\end{equation}
It can be seen that this bound is on each and every elements 
of the mass matrix and not on the physical states and 
independent on the existence of any right handed neutrinos.

In general, a Majorana particle can be described by a four component real
field,
\begin{equation}
\Psi_M = \sqrt{m_\nu \over E_\nu} [u_\nu (b_\nu + d_\nu^*) 
{\rm e}^{(-ip.x)} + v_\nu (b_\nu^* + d_\nu) {\rm e}^{(ip.x)} ]
\end{equation}
and hence the charged  current  containing  a Majorana  field, $$
j_\mu =  \overline{\Psi}_l  \gamma_\mu  (1 - \gamma_5)  \Psi_M $$
will have a lepton number  violating  part.  However, this lepton
number  violating  contribution  will always be  suppressed  by a
factor  $(m_\nu/E_\nu)$ and hence the rate of such processes will
be suppressed by a factor $(m_\nu/E_\nu)^2$.  Thus even the decay
of the $W^\pm$ to $e$ and $\nu$ will have lepton number violation
at a rate,
$$ \Gamma (W \rightarrow e \nu) = \frac{\alpha_W}{4} \frac{m_\nu^2 
M_W^2}{T^2 (T^2 + M_W^2)^{1/2}} . $$
The survival of baryon asymmetry of the universe after the electroweak
phase transition again requires this process to be alow enough,
$$ \Gamma(W \to e \nu) < H . $$ This translates to a bound on the Majorana 
mass of the left handed neutrino,
\begin{equation}
m_\nu < 30 \hskip .2in {\rm eV} .
\end{equation}

Similarly,  the decay of the higgs  doublet to an electron and an
anti-neutrino    will   also   have   lepton   number   violating
contribution,  but they will be suppressed by the Yukawa coupling
constants and cannot give stronger bounds.  Similarly  scattering
processes  involving  the higgs, like $\phi + \phi \to l_i + l_j$
(mediated by a virtual  left-handed  neutrino) will contribute to
the evolution of the lepton number  asymmtry of the universe, but
it will  be  much  suppressed  compared  to the  charged  current
interactions and hence cannot give stronger bound to the Majorana
mass of the left-handed neutrinos.

To summarize I point out that it is possible to constraint the
Majorana masses of the left-handed neutrinos directly from the
survival of baryogenesis after the electroweak phase transition.
The strongest bound comes from the process $W + W \to e + e$ (as
shown in figure 1).
This bound of about 20 keV  does not depend on 
the details of the models and may not be avoided even if the 
particles are not stable. 

{\bf  Acknowledgement}  I would  like  to  thank  Professor  E.A.
Paschos, Univ Dortmund, Germany for hospitality, where part of this
work has been done and Alexander von Humboldt Foundation for a 
fellowship.


\begin{figure}[htb]
\mbox{}
\vskip 4.5in\relax\noindent\hskip -.6in\relax
\includegraphics{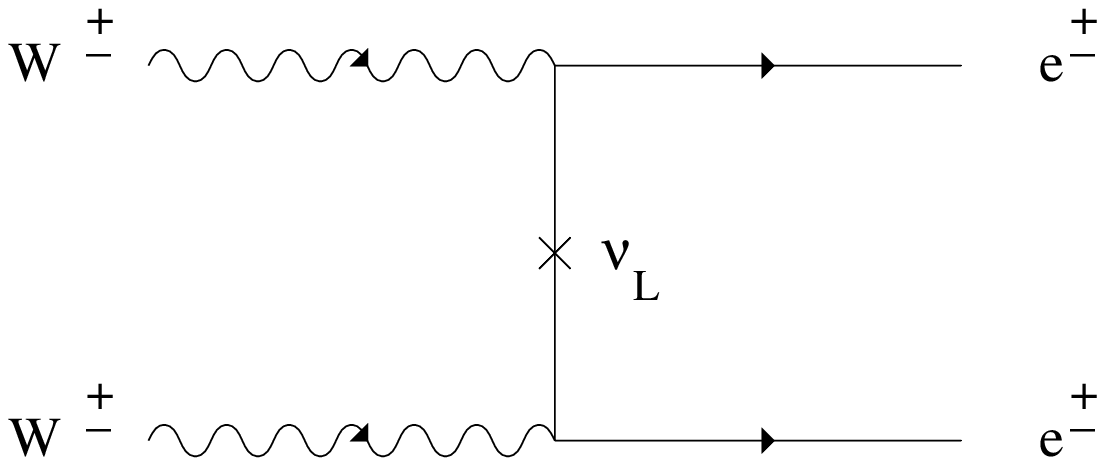}
\caption{ Lepton number violating processes $W^{\pm} + 
W^{\pm} \to e^{\pm} + e^{\pm}$ mediated by the left handed Majorana 
neutrinos.}
\end{figure}
\end{document}